\documentclass[10pt, a4paper]{article}

\usepackage[]{lrec-coling2024} 
\usepackage{booktabs}
\usepackage{amsmath}
\usepackage[absolute,overlay]{textpos}
\newtheorem{definition}{Definition}

\title{Large Language Models for Generative Recommendation: A Survey and Visionary Discussions}

\name{Lei Li$^1$, Yongfeng Zhang$^2$, Dugang Liu$^3$, Li Chen$^1$} 

\address{$^1$Hong Kong Baptist University, Hong Kong, China\\
	$^2$Rutgers University, New Brunswick, USA\\
	$^3$Guangdong Laboratory of Artificial Intelligence and Digital Economy (SZ), Shenzhen, China\\
         \{csleili, lichen\}@comp.hkbu.edu.hk,
         yongfeng.zhang@rutgers.edu,
         dugang.ldg@gmail.com\\}

\abstract{
Large language models (LLM) not only have revolutionized the field of natural language processing (NLP) but also have the potential to reshape many other fields, e.g., recommender systems (RS). However, most of the related work treats an LLM as a component of the conventional recommendation pipeline (e.g., as a feature extractor), which may not be able to fully leverage the generative power of LLM. Instead of separating the recommendation process into multiple stages, such as score computation and re-ranking, this process can be simplified to one stage with LLM: directly generating recommendations from the complete pool of items. This survey reviews the progress, methods, and future directions of LLM-based generative recommendation by examining three questions: 1) \textit{What} generative recommendation is, 2) \textit{Why} RS should advance to generative recommendation, and 3) \textit{How} to implement LLM-based generative recommendation for various RS tasks. We hope that this survey can provide the context and guidance needed to explore this interesting and emerging topic.
 \\ \newline }

\begin{document}
\begin{textblock*}{10cm}(2cm,1cm) 
	\underline{Published as a conference paper at LREC-COLING 2024}
\end{textblock*}
\maketitleabstract

\section{Introduction}

Large language models (LLM) are profoundly affecting the field of natural language processing (NLP), and their powerful ability to handle various tasks has also inspired new paths for practitioners in other fields, e.g., recommender systems (RS).
As an effective means to solve information overload in today's era, RS has been closely integrated into our daily lives, and how to effectively reshape it with LLM is a promising research issue \cite{P5}.
Although natural language is an expressive medium, it can also be vague.
For example, when an LLM is deployed for vehicle identification and scheduling, using vague descriptions (e.g., ``a black SUV'') to identify a vehicle rather than a precise identifier such as vehicle identification number (VIN) or plate number would be dangerous. 
Similarly, vagueness could also be a problem in recommendation scenarios that require precise and unique identifiers of items, because RS needs to guarantee that recommendations made for users are things that factually exist to avoid the hallucination problem \cite{azamfirei2023large}.
This also explains why an ID is usually assigned for each user/item in RS.
Despite that, the current understanding of IDs is usually limited to one form, i.e., most RS research considers each ID as a discrete token associated with an embedding vector.
In this survey, we generalize the definition of ID to strengthen its connection with LLM:
\begin{definition}[ID in Recommender Systems]
	An ID in recommender systems is a sequence of tokens that can uniquely identify an entity, such as a user or an item.
	An ID can take various forms, such as an embedding ID, a sequence of numerical tokens, and a sequence of word tokens (including an item title, a description of the item, or even a complete news article), as long as it can uniquely identify the entity.
	\label{def:id}
\end{definition}

For example, a product on an e-commerce platform may be assigned the ID \textit{item\_7391} and be further represented as a sequence of tokens such as $\langle$item$\rangle$$\langle$\_$\rangle$$\langle$73$\rangle$$\langle$91$\rangle$ \cite{P5,OpenP5}.
Note that the ID may not necessarily be comprised of numerical tokens.
As long as it is a unique identifier for an item, it can be considered as the item's ID.
For instance, the title of the movie ``The Lord of the Rings'' can be considered as this movie's ID.
The ID could even be a sequence of words that do not convey any explicit meaning, e.g., ``ring epic journey fellowship adventure'' \cite{Index}.
IDs in conventional RS can be seen as a special case of the above definition, i.e., a sequence of just one token.
Under this definition, IDs resemble token sequences as in text and thus naturally fit the natural language environment and LLM.

Due to the huge number of items in real-world systems, traditional RS usually takes the multi-stage filtering paradigm \cite{covington2016deep} -- some simple and efficient methods (e.g., rule-based filtering) are used to reduce the number of candidate items from millions to a few hundred, and advanced recommendation algorithms are then applied on these items to further select a few number of items for recommendation.
As a result, advanced recommendation algorithms are not applied to all items, but only to a few hundred items.

The generative power of LLM has the potential to reshape the RS paradigm from multi-stage filtering to single-stage filtering.
Specifically, an LLM itself can be the single and entire recommendation pipeline that directly generates the items for recommendation, eliminating the need for multi-stage filtering.
In this way, advanced LLM-based recommendation algorithms are applied to all items in the system but in an implicit manner.
We term such a process \textit{generative recommendation} and formally define it as follows:
\begin{definition}[Generative Recommendation]
	A generative recommender system directly generates recommendations or recommendation-related content without the need to calculate each candidate's ranking score one by one.
\end{definition}
\noindent In a broader sense, this is in line with the trend of general artificial intelligence (AI) research, which recently has been shifting from discriminative AI (such as classification and regression) to generative AI (e.g., ChatGPT\footnote{\url{https://openai.com/chatgpt}}).

With the above definitions, we first answer why RS is developing towards generative recommendation in Section \ref{sec:why}.
In Section \ref{sec:id}, we review ID creation approaches that could retain the collaborative information of IDs in the LLM environment.
Then, we show how typical recommendation tasks can be performed with LLM by providing general formulation in Section \ref{sec:how}, and highlight opportunities in the LLM era in Section \ref{sec:future}.
At last, we conclude our survey in Section \ref{sec:conclude}.

It should be noted that our survey is different from some recent surveys on LLM-based recommendation \cite{training,wu2023survey,fan2023recommender,Industry,chen2023large,vats2024exploring,huang2024foundation} from two perspectives: 1) our survey is organized with generative recommendation as the key focus, eliminating discriminative recommendation models for clarity; 2) we develop a taxonomy for LLM-based recommendation research with strong inspiration from the recommendation community, instead of blindly following the LLM taxonomy from NLP community.

To sum up, this survey makes the following contributions:
\begin{itemize}
	\item To the best of our knowledge, this is the first survey that systematically summarizes research on LLM-based generative recommendation. To differentiate this topic from traditional RS, we have generalized the definition of ID for generative recommendation.
	\item This survey is pragmatic as we provide the formulation for different LLM-based recommendation tasks when categorizing relevant research, which would provide a useful guideline for future research.
	\item We discuss important and promising directions to explore for LLM-based generative recommendation research, which may help broaden the scope of this under-explored research area.
\end{itemize}

\begin{figure*}[htbp]
	\centering
	\includegraphics[scale=0.45]{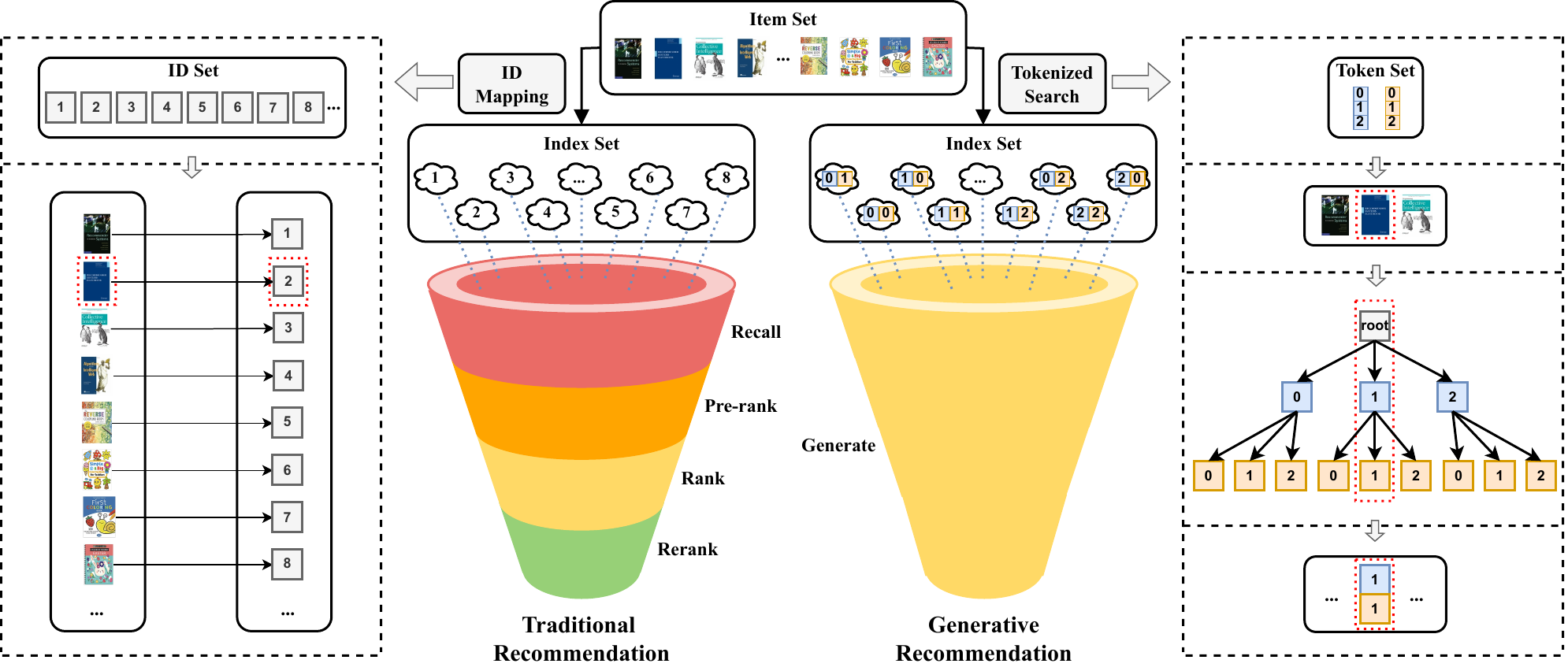}
	\caption{Pipeline comparison between traditional recommender systems and LLM-based generative recommendation.}
	\label{fig:pipeline}
\end{figure*}

\section{Why Generative Recommendation} \label{sec:why}
To answer why RS is developing towards generative recommendation, we first discuss problems with discriminative recommendation.
When the number of items on a recommendation platform is prohibitively large, calculating the ranking score for each item would be computationally expensive.
Therefore, industrial RS usually consists of multiple stages to narrow down the candidate items.
At the early stage, simple models (e.g., logistic regression) or straightforward filtering strategies (e.g., feature matching) are usually adopted to filter out less relevant items.
Only in the final stage can the relatively complex and advanced models be utilized.
This naturally causes a gap between academic research and industrial applications.
Although recent recommendation models are growing more fancy and sophisticated, few have been practically employed in industry.

In the era of LLM, we see a great opportunity that could potentially bridge this gap.
As both academic research and industry applications may share the same backbone LLM, most research advancements on LLM could benefit its downstream applications.
Regarding the recommendation pipeline, the typical multiple stages could be advanced to one stage for generative recommendation, i.e., directly generating items for recommendation.
A graphical comparison between the two types of pipeline is shown in Fig. \ref{fig:pipeline}.
At each step of recommendation generation, an LLM can produce a vector that represents the probability distribution on all possible ID tokens.
After a few steps, the generated tokens can constitute a complete ID that stands for the target item.
This process implicitly enumerates all candidate items for generating the target item for recommendation, which differs from traditional RS, which draws items from a subset resulted from the previous filtering step.

The key secret of LLM for generative recommendation is that we can use finite tokens to represent almost infinite items.
Suppose that we have $1000$ tokens for representing user or item IDs, which can be numerical tokens, word tokens, or even out-of-vocabulary (OOV) tokens, and each ID consists of $10$ tokens, then we can use these $1000$ tokens to represent as many as $1000^{10} = 10^{30}$ items (i.e., unique IDs), which is almost an astronomical number and large enough for most of real-world RS. 
When applying the beam search algorithm \cite{beam} for generating item IDs, the probability vector at each step is bounded by $1000$ tokens, making it computationally possible to directly generate recommendations out of the item pool.

\section{ID Creation Methods} \label{sec:id}

When implementing generative recommendation with LLM, the input (particularly user and item IDs) should be made into the right format that is compatible with LLM.
Intuitively, one would consider the metadata of users and items as an alternative, such as user name and item title.
This type of ID representation is quite common in related work, as summarized in Table \ref{tab:id}.
Despite the popularity, it has two problems \cite{li2023exploring}.
First, when the IDs are extremely long, e.g., in the case of item description, it would be computationally expensive to conduct generation.
Besides, it would be difficult to find an exact match in the database for a long ID;
and double-checking the existence of each ID would take us back to discriminative recommendation since we need to compare it with each item in the database.
Second, although natural language is a powerful and expressive medium, it can also be vague in many cases.
For example, two irrelevant items could have identical names, such as Apple the fruit and Apple the company, while two closely related items may have distinct titles, as in the well-known ``beer and diaper'' example in data mining.

\begin{table*}[htbp]
	\tiny
	\centering
	\begin{tabular}{p{1.3cm}p{1.6cm}p{11.5cm}}
		\toprule
		\textbf{Item ID}&\textbf{User ID}&\textbf{Related Work}\\
		\cmidrule(lr){1-3}
		
		Token Sequence (e.g., ``\textit{56 78}'') & Token Sequence & P5 \cite{P5}, UP5 \cite{UP5}, VIP5 \cite{VIP5}, OpenP5 \cite{OpenP5}, POD \cite{POD}, GPTRec \cite{SVD}, TransRec \cite{TransRec}, LC-Rec \cite{LC-Rec}, \cite{Index}\\
		\cmidrule(lr){1-3}
		
		Item Title (e.g., ``\textit{Dune}'')&Interaction History (e.g., [``\textit{Dune}'', ``\textit{Her}'', ...]) &LMRecSys \cite{LMRecSys}, GenRec \cite{GenRec}, TALLRec \cite{TALLRec}, NIR \cite{NIR}, PALR \cite{PALR}, BookGPT \cite{BookGPT}, PBNR \cite{PBNR}, ReLLa \cite{ReLLa}, BIGRec \cite{BIGRec}, TransRec \cite{TransRec}, LLaRa \cite{LLaRa}, Llama4Rec \cite{Llama4Rec},  Logic-Scaffolding \cite{Logic-Scaffolding}, \cite{dai2023uncovering,liu2023chatgpt,hou2023large,News,zhang2023recommendation,wang2023rethinking,lin2023sparks,di2023evaluating,li2023exploringnews}\\
		\cmidrule(lr){1-3}
		
		Item Title&Metadata (e.g., age)&InteRecAgent \cite{InteRecAgent}, \cite{zhang2023chatgpt,he2023large}\\
		\cmidrule(lr){1-3}
		
		Metadata&Metadata&M6-Rec \cite{M6}, LLMRec \cite{LLMRec}, RecMind \cite{RecMind}, TransRec \cite{TransRec}, \cite{wu2023exploring}\\
		\cmidrule(lr){1-3}
		
		Embedding ID&Embedding ID&PEPLER \cite{PEPLER}\\
		\bottomrule
	\end{tabular}
	\caption{Methods of representing IDs for LLM-based generative recommendation.}
	\label{tab:id}
\end{table*}

Therefore, we need concise and unique representations of IDs in recommendation scenarios to precisely distinguish one user or item from the others.
Associating each ID with an embedding vector is a common practice in traditional RS, but it would cost a lot of memory to store them since industry-scale RS usually involves tons of users and items.
In addition, these IDs are OOV tokens to LLM and thus are not very compatible with them.

This is why a new way of representing IDs, i.e., a sequence of tokens rather than a single embedding, is needed.
The key idea is to use a small amount of tokens to represent an astronomical number of users or items, as explained in the previous section.
To make IDs reasonably short, similar users or items could share more tokens in their ID sequences, while the remaining tokens can be used to guarantee their uniqueness.
In the following, we review three typical ID creation approaches that follow this principle.
Most of these ID creation methods aim to encode the user-user, item-item, or user-item collaborative information into IDs, which combines the merit of collaborative filtering from traditional RS with the emerging LLM for effective recommendation.

\subsection{Singular Value Decomposition}

\cite{SVD} acquire an item's ID tokens from its latent factors.
Specifically, they first perform truncated singular value decomposition on user-item interaction data to obtain the item embedding matrix.
After a set of operations, including normalization, noise-adding, quantization, and offset adjustment, each item's embedding becomes an array of integers, which serves as this item's ID sequence.
In particular, the noise-adding operation can ensure that there are no identical item embeddings, and thus make each item ID unique.

\subsection{Collaborative Indexing}

\cite{Index} compose an item ID with nodes on a hierarchical tree.
Technically, they first construct an item graph whose edge weights denote the co-occurring frequency of any two items in all users' interaction history.
Then, the graph's adjacency matrix and Laplacian matrix, as well as the latter's eigenvectors, can be computed.
With the eigenvectors, spectral clustering \cite{Spectral} can be applied to group similar items into the same cluster.
By recursively doing so, large clusters can be further divided into smaller ones.
When the number of nodes in each cluster is smaller than a threshold, these clusters and their sub-clusters naturally constitute a hierarchical tree whose leaf nodes are the items.
After assigning tokens to each node, each item has a unique ID sequence by following a path from the root node to the leaf node.

\subsection{Residual-Quantized Variational AutoEncoder}

\cite{LC-Rec} quantize item embeddings with residual-quantized variational auto-encoder (RQ-VAE) \cite{RQVAE} to obtain item IDs.
They first encode an item's textual description with an LLM to get the item's embedding.
After passing the embedding through VAE's encoder, a latent representation can be acquired.
Then, they treat this representation as the initial residual vector and perform multi-step residual quantization.
At each step, there is a codebook, i.e., an embedding table, from which the nearest embedding to the residual vector can be found.
The index of this embedding in the codebook will be the item's codeword at this step, i.e., a token of the item ID sequence.
A new residual vector can be computed by subtracting the old residual vector with the nearest embedding.
By repeatedly doing so, the complete item ID can be formed.

In addition to the above three ID creation approaches, \cite{Index} discussed other strategies, such as sequential indexing based on user interaction history and semantic indexing based on item metadata information, which are effective approaches to creating item IDs.
We omit the details because they are quite simple.

\section{How to Do Generative Recommendation} \label{sec:how}

With the above-defined user and item IDs, we now describe how to perform different generative recommendation tasks with LLM.
A summary of relevant research on each task is given in Table \ref{tab:task}.
We can see that there are a few models that can perform multiple recommendation tasks, e.g., P5 \cite{P5}.
To allow LLM to understand each task, especially those that have the same input data, we can construct a prompt template \cite{Prompt} that describes the task and then fill the user and item information, such as their IDs, in the prompt.
During the inference stage, all sorts of output (e.g., predicted item IDs) are generated auto-regressively in the way of natural language generation.
Next, we introduce the general formulation of each task, followed by the recent progress.
Finally, we discuss how to evaluate these tasks.

\begin{table*}[htbp]
	\tiny
	\centering
	\begin{tabular}{p{1.4cm}p{2.8cm}p{3.3cm}p{1.8cm}p{1cm}p{1.1cm}p{1.5cm}}
		\toprule
		\textbf{Rating Prediction} & \textbf{Top-N Recommendation} & \textbf{Sequential Recommendation} & \textbf{Explainable Recommendation} & \textbf{Review Generation} & \textbf{Review Summarization} & \textbf{Conversational Recommendation}\\
		\midrule
		P5 \cite{P5}, BookGPT \cite{BookGPT}, LLMRec \cite{LLMRec}, RecMind \cite{RecMind}, Llama4Rec \cite{Llama4Rec}, \cite{liu2023chatgpt,dai2023uncovering,li2023exploringnews}
		& P5 \cite{P5}, UP5 \cite{UP5}, VIP5 \cite{VIP5}, OpenP5 \cite{OpenP5}, POD \cite{POD}, GPTRec \cite{SVD}, LLMRec \cite{LLMRec}, RecMind \cite{RecMind}, NIR \cite{NIR}, Llama4Rec \cite{Llama4Rec}, \cite{zhang2023chatgpt,zhang2023recommendation,liu2023chatgpt,News,dai2023uncovering,di2023evaluating,carraro2024enhancing}
		& P5 \cite{P5}, UP5 \cite{UP5}, VIP5 \cite{VIP5}, OpenP5 \cite{OpenP5}, POD \cite{POD}, GenRec \cite{GenRec}, GPTRec \cite{SVD}, LMRecSys \cite{LMRecSys}, PALR \cite{PALR}, LLMRec \cite{LLMRec}, RecMind \cite{RecMind}, BIGRec \cite{BIGRec}, TransRec \cite{TransRec}, LC-Rec \cite{LC-Rec}, LLaRa \cite{LLaRa}, \cite{Index,liu2023chatgpt,hou2023large,zhang2023recommendation}
		& P5 \cite{P5}, VIP5 \cite{VIP5}, POD \cite{POD}, PEPLER \cite{PEPLER}, M6-Rec \cite{M6}, LLMRec \cite{LLMRec}, RecMind \cite{RecMind}, Logic-Scaffolding \cite{Logic-Scaffolding}, \cite{liu2023chatgpt}
		& -
		& P5 \cite{P5}, LLMRec \cite{LLMRec}, RecMind \cite{RecMind}, \cite{liu2023chatgpt}
		& M6-Rec \cite{M6}, RecLLM \cite{RecLLM}, InteRecAgent \cite{InteRecAgent}, PECRS \cite{PECRS}, \cite{wang2023rethinking,lin2023sparks,he2023large}\\
		\bottomrule
	\end{tabular}
	\caption{Seven typical generative recommendation tasks with LLM.}
	\label{tab:task}
\end{table*}

\subsection{Rating Prediction}
In conventional RS, the rating prediction task is formulated as follows: given a user $u$ and an item $i$, a recommendation model $f(u, i)$ needs to estimate a score $\hat{r}_{u, i}$ that the user would rate the item.
In the context of LLM, $u$ and $i$ are no longer embedding IDs, but two sequences of tokens as defined in Definition \ref{def:id}.
The two IDs can be filled in an instruction prompt $p(u, i)$, e.g., ``how would \textit{user\_1234} rate \textit{item\_5678}'', such that LLM can understand this task.
After feeding $p(u, i)$ into an LLM, it can generate a numerical string on a scale of 1 to 5, such as ``4.12'', indicating that the user is likely to interact with the item.

There are some studies \cite{P5,Llama4Rec} that tested LLM with this task, among which many \cite{BookGPT,dai2023uncovering,liu2023chatgpt,LLMRec,RecMind,li2023exploringnews} are based on ChatGPT.
As users may not leave an explicit rating for each item with which they interacted, the rating prediction task can be less practical for real-world systems.
Instead, implicit feedback, e.g., clicking, is easier to collect.
Thus, how to infer users' preferences from such implicit feedback motivates the development of the top-$N$ recommendation task.

\subsection{Top-$N$ Recommendation}

The top-$N$ recommendation task, a.k.a., ranking, aims to select $N$ items as recommendations for a given user $u$.
To this end, traditional RS usually computes a score $\hat{r}_{u, i}$ w.r.t. each item $i$ in the item set $\mathcal{I}$.
After filtering out those that the user already interacted with, i.e., $\mathcal{I}_u$, the top candidates can be selected as recommendations from the remaining items as $\text{Top}(u, i) := \mathop{\arg\max}_{i \in \mathcal{I}/\mathcal{I}_u}^{N} \hat{r}_{u, i}$.

However, due to the context length limit of an LLM, it is impossible to feed the model all the items.
As a result, the community has explored two approaches to tackle the problem.
One is \textit{straightforward recommendation} \cite{OpenP5,zhang2023chatgpt,di2023evaluating}, which uses a prompt that only contains a user's information (e.g., ID or metadata) and asks the LLM to directly generate recommendations for this user.
The second is \textit{selective recommendation} \cite{P5,VIP5,POD,UP5,SVD,zhang2023recommendation,News,dai2023uncovering,liu2023chatgpt,LLMRec,RecMind,NIR,Llama4Rec,carraro2024enhancing}, which provides both user information and a list of candidate items $\mathcal{I}_c$ in the prompt and asks the LLM to select items for recommendation out of these candidates.
The candidate list could be comprised of a testing item and several sampled negative items.
After filling the user and candidates in a prompt $p(u, \mathcal{I}_c)$, e.g., ``select one item to recommend for \textit{user\_1234} from the following candidates: \textit{item\_6783, ..., item\_9312, item\_2834}'', the LLM can generate an item ID (e.g., ``9312'') as recommendation.
When combined with beam search, the model can produce several item IDs and thus a list of $N$ recommendations.

Besides generating item IDs, some recent studies \cite{PBNR} instruct LLM to answer whether a user is going to interact with a given item by generating ``yes'' or ``no''.
Although the ``yes'' or ``no'' answer is generated by LLM, these methods can be considered as discriminative recommendations since they need to generate an answer or a score (e.g., the probability of ``yes'') for each item.

\subsection{Sequential Recommendation}

The sequential recommendation task goes one step further than the top-$N$ recommendation with the consideration of the time or order of interaction.
Specifically, its objective is to predict the next item with which a user $u$ is likely to interact based on his/her past interactions.
The items interacted by the user are chronologically ordered according to their timestamps, which can be denoted as $I_u$.
Considering the sequential nature of such data, researchers usually employ sequential models to deal with the problem, such as Markov chains, recurrent neural networks (RNN), and Transformer \cite{Transformer}.
Again, we can first fill the user and the item sequence in a prompt $p(u, I_u)$, e.g., ``given \textit{user\_1234}'s interaction history \textit{item\_3456, ..., item\_4567, item\_5678}, predict the next item with which the user will interact'', and then prompt LLM to generate an item ID as a prediction, e.g., ``6789''.
To reduce the inference time, we can cut off the relatively old items before filling the item sequence in the prompt.

This task is a trending problem, as evidenced by a significant number of LLM-based models \cite{P5,VIP5,OpenP5,POD,SVD,LMRecSys,UP5,Index,liu2023chatgpt,LLMRec,zhang2023recommendation,RecMind,LC-Rec}.
In particular, \cite{BIGRec,TransRec} leverage LLM to generate candidates for further filtering while \cite{PALR,hou2023large,GenRec,LLaRa,RecRanker} provide LLM with candidate items for recommendation, and \cite{TALLRec,ReLLa,zhang2023recommendation} instruct LLM to answer whether a user will like a specific item.

\subsection{Explainable Recommendation}

Besides generating recommendations, explanations that allow users to know the reason behind them are equally important.
There are various methods to explain a recommendation to a user, such as explicit item features \cite{EFM} and visual highlights \cite{VECF}.
We refer interested readers to the survey \cite{Explanation} for a comprehensive examination of explainable recommendations.

A typical LLM-based recommendation explanation task can be natural language explanation generation.
That is, given a pair of user $u$ and item $i$, we direct the model to generate a sentence or paragraph in natural language to explain why $i$ is recommended to $u$.
Ground-truth explanations can be mined from user reviews \cite{NETE}.
As the inputs (i.e., $u$ and $i$) are identical to those for rating prediction, we can put them in a prompt $p(u, i)$ to inform the LLM that this is an explanation task, e.g., ``explain to \textit{user\_1234} why \textit{item\_5678} is recommended.''
As a response, the model may generate an explanation such as ``The movie is top-notch.''
However, using IDs alone in the prompt may be unclear as to which aspects the model should discuss in the explanation.
To address this problem, we can provide some item features $f$ as hint words in the prompt, e.g., ``acting''.
An example prompt $p(u, i, f)$ for such a scenario could be ``write an explanation for  \textit{user\_1234} about \textit{item\_5678} on the feature \textit{acting}.''
Then, the LLM may generate an explanation such as ``The acting in this movie is attractive.''

\cite{P5,VIP5,POD,liu2023chatgpt,LLMRec,RecMind} perform the explanation generation task as above;
\cite{M6} trigger the explanation task with the keyword ``because'';
\cite{Logic-Scaffolding} adopt chain-of-thought prompting with multiple steps of reasoning;
\cite{PEPLER} make use of continuous prompt vectors instead of discrete prompt templates.

\subsection{Review Generation}

In addition to explanation generation, the above formulation can also be adapted to the review generation task \cite{Review}, which may make it easier and more efficient for users to leave a comment after purchasing a product, watching a movie, taking a ride, etc. 
The resulting data would in turn facilitate the development of recommendation-related research, such as explainable recommendations and conversational recommendations.
As usual, we can fill a user and his/her interacted item in a prompt $p(u, i)$, e.g., ``generate a review for \textit{user\_1234} about \textit{item\_5678}.''
Then, the LLM may generate a review paragraph. For example, ``the hotel is located in ...''.
However, we have not noticed any LLM-based recommendation research on this problem, probably because the formulation is too similar to explanation generation, except that reviews are generally longer.

\subsection{Review Summarization}

Reading a long review can take some time, which users may not always be able to afford.
A highly concise review summary can help users quickly understand the pros and cons of an item.
Current LLM-based review summarization methods \cite{P5,liu2023chatgpt,LLMRec,RecMind} mainly target how to summarize a user $u$'s own review $R$ for an item $i$, and treat the review title or tip as the summary.
In this case, we can construct a prompt and fill the ternary data in $p(u, i, R)$, e.g., ``summarize the following review that \textit{user\_1234} wrote for \textit{item\_5678}: \textit{the hotel is located in ...}''.
Then, the LLM may generate a summary, e.g., ``great location''.

However, it may be unnecessary to summarize a user's review because the user already knows about the reviewed item.
Instead, summarizing the review for another user who has never interacted with the item can be more useful.
Furthermore, it is also meaningful to conduct a multi-review summarization that summarizes different users' opinions on the same item.

\subsection{Conversational Recommendation}

The goal of conversational recommendation is to recommend a user some items within multiple rounds of conversation \cite{Conversation,sun2018conversational,zhang2018towards}.
Different from traditional RS which mainly relies on users' historical interactions, in a conversational environment users can freely state their preferences in natural language and even provide negative feedback, e.g., rejecting a recommendation.
However, the research community is still in the process of reaching a consensus on how to formulate this task.

\cite{M6,RecLLM,he2023large} adopt two labels (i.e., ``USER'' and ``SYSTEM'') to mark the speaker of an utterance before feeding a dialogue session into LLM for generating a response.
\cite{InteRecAgent} instruct LLM to call tools, such as traditional recommenders and SQL, to narrow down candidate items for recommendation.
\cite{lin2023sparks} directly chat with ChatGPT, because they aim to establish principles for conversational recommendation, e.g., memory mechanism and repair mechanism, rather than developing new models.
For evaluation, \cite{wang2023rethinking} point out the problem of current protocols.
Specifically, although ChatGPT's chatting ability is undeniably impressive, its performance on existing metrics is not very good because they overly stress the matching between generated responses and annotated recommendations or utterances.

\subsection{Evaluation Protocols}

To evaluate the performance of LLM on these tasks, we can apply existing metrics.
For rating prediction, root mean square error (RMSE) and mean absolute error (MAE) are commonly used.
For the other two recommendation tasks, i.e., top-$N$ recommendation and sequential recommendation, we can employ ranking-oriented metrics, such as normalized discounted cumulative gain (NDCG), precision, and recall.
Besides offline evaluation, online A/B tests can also be adopted since they can reflect users' actual interactions with recommended items.

As to natural language generation tasks, including explanation generation, review generation, review summarization, and conversational recommendation, the quality of LLM's generation can be estimated with BLEU \cite{BLEU} in machine translation and ROUGE \cite{ROUGE} in text summarization.
Both metrics measure the degree of matching between text segments of the generated content and those of the ground-truth. 
However, as pointed out by \cite{wang2023rethinking}, it can be problematic to over-emphasize the matching with annotated data.
Also, there are other aspects beyond text similarity that cannot be reflected by BLEU or ROUGE.
As an early attempt, \cite{NETE} proposed several metrics such as feature coverage ratio and feature diversity that take into account the characteristics of explicit elements for the evaluation of explanations, but they are still rudimentary.
Although there are some other learning-based metrics, e.g., BERTScore \cite{zhang2019bertscore}, more advanced and standard metrics need to be developed.
In addition to automatic evaluation, we can also conduct human evaluation to measure LLM on these generation tasks.
However, it requires researchers to properly design the questionnaire and the number of participants could be limited. 

\section{Challenges and Opportunities}  \label{sec:future}
In this section, we discuss research challenges and opportunities for generative recommendation in the LLM era, especially those significant matters that need urgent care.

\subsection{LLM-based Agents}

Simulators have played an important role in addressing the data-scarcity problem in RS, especially in conversational recommendation environments where the ground-truth interaction data are usually unavailable \cite{yu2023counterfactual}.
Recently, we have seen that LLM-based generative agents could simulate almost any scenario, such as a small society \cite{agent} or world wars \cite{hua2023war}.
There also emerges user behavior simulators for RS \cite{RecAgent,Agent4Rec}.
However, a paradox arises when applying simulators to RS.
On one hand, if the interaction data simulated by a simulator do not align with the target user's true preference, then the simulated data may not be truly useful.
On the other hand, if the simulator can perfectly simulate a user's preference, then we may not need recommendation algorithms at all since the simulated data can be directly adopted as recommendations.

We believe that the potential of LLM-based agents for RS is beyond simple simulation.
Nowadays, they can call tools, APIs, and expert models to solve complex tasks that take several steps of reasoning \cite{ge2024openagi}.
Such an ability could push LLM-based RS to a broader range of real-world applications.
Taking trip recommendation as an example, the system can cater to a user's personalized needs, such as duration, budget, and preferred attractions, and draft an itinerary by looking up real-time information, such as the attractions' opening hours and the transportation time from one attraction to another.
When such a system is embedded in vehicles \cite{vehicle}, it can even route for drivers by calling map APIs, and also recommend out-of-vehicle services, such as restaurants and charging/gas stations.
No matter what scenario, sometimes users may not be able to follow the itinerary, and in this case the system can dynamically revise it to fit the user's current status.
By connecting with real-world objects, this new generation of RS has the potential to change how people live.

\subsection{Hallucination}

Hallucination \cite{azamfirei2023large} means that the content generated by an LLM may deviate from facts. Hallucination is an important problem in LLM as well as their applications. 
In particular, for LLM-based RS, we need to guarantee that the items recommended to users exist, otherwise it may cause user dissatisfaction and frustration and even low user adoption of the system in real life. 
For example, a user may spend time traveling to a recommended restaurant, only to find out that such a restaurant does not exist at all. 
Particularly, in high-stake recommendation domains such as drug recommendation, medical treatment recommendation, and financial investment recommendation, hallucinated recommendations may cause severe losses for users. 

There are two possible approaches to addressing the hallucination problem in LLM-based RS.
One is to use meticulously designed item IDs for generation.
For example, \cite{Index} create item IDs and organize them into a prefix tree structure, which is also called a trie structure. 
As long as the beam search generation process follows the root-to-leaf paths in the tree, the generated items will always exist. 
The other method is to apply retrieval-augmentation over an LLM \cite{mialon2023augmented}, i.e., conditioning an LLM on retrieved items, so that the recommended items match those in the item database.
Furthermore, the two methods, i.e., indexing and retrieval, can be integrated to address the hallucination problem effectively and efficiently. 

\subsection{Bias and Fairness}

There can be two types of bias for LLM-based RS, which are \textit{content bias} and \textit{recommendation bias}.
The former refers to the bias that can be directly observed in the generated content.
A typical example is gender bias.
\cite{COFFEE} find that machine-generated recommendation explanations for male users are usually longer than those for female users in the game domain.
This problem may lie in the training data that are adapted from user reviews of games.
In addition, an LLM trained with a huge amount of human-generated data may reiterate or even reinforce the bias hidden in the data.
Taking linguistic bias as an example, \cite{LMRecSys} observe that LLM tend to use generic tokens when generating item titles to make them look fluent and linguistically sound, but lead to recommendations that are greatly different from users' preferred items.
When adapted to downstream recommendation tasks, the bias should be mitigated or even completely removed to prevent the propagation of negative effects and to improve user experience.

Regarding recommendation bias, \cite{News} report that ChatGPT is prone to recommend news articles from providers that it labeled as popular.
\cite{xu2023llms} observe domain difference when asking ChatGPT to recommend news articles and jobs for varying gender identities and races.
Similarly, \cite{zhang2023chatgpt} find that the music recommendations made by ChatGPT for people with different demographic attributes (e.g., white v.s. African American) are dissimilar.
Although the results look biased, they could also be a type of personalization since the music tastes of people under different cultural backgrounds could differ.
Therefore, a question needs to be answered: \textit{What is the boundary between bias and personalization?}
\cite{UP5} attempt to make LLM-based recommendation models fair concerning sensitive attributes, such as age, marital status, and occupation, by distilling the bias into continuous prompts.
As the bias and fairness issues are still open problems, more work should be done, e.g., from the perspective of fairness definition and bias mitigation for LLM-based RS.

\subsection{Transparency and Explainability}

Making recommendations transparent and explainable to users has always been an important problem for RS and AI in general \cite{Explanation}.
Due to the huge size and complexity of LLM, explaining LLM-based recommendations has posed new challenges to the community.
There are two types of explainability for LLM-based RS.
One is to generate reasonable natural language explanations for recommended items, while the other is to dive into the model and try to explain the internal working mechanism of an LLM.
While researchers have explored the first type of explainability for a while \cite{PETER,PEPLER,POD,P5,VIP5,M6}, the second type of explainability has been largely unexplored.
One possible method is to align an LLM such as its prompts with an explicit knowledge base such as a knowledge graph \cite{geng2022path,ye2023natural} so that the model's decision-making process is aligned with explicit paths in the knowledge graph for explanation.
However, the direction is generally very preliminary and requires innovative methods and brave new ideas from the community.

\subsection{Controllability}

Controllability is an important problem for LLM since we usually cannot precisely control the output of an LLM.
The lack of controllability may cause serious problems.
For example, an LLM may generate harassing content, fake content, or content that deviates from basic moral standards.
For RS, the controllability issue is more complicated due to the various recommendation tasks or scenarios that require controllability \cite{tan2023user,wang2022user,schafer2002meta,parra2015user}.
For example, users may want to control the feature that an explanation talks about \cite{PETER,NETE,P5}, e.g., if a user cares about the ``price'' of a restaurant, then the explanation should talk about its price, while if the user is concerned about ``distance'', then the explanation should discuss the distance.
Besides the controllability of explanations, users may also want to control the features of recommended items, such as price level, color, and brand \cite{tan2023user}.
For example, users may hope that an LLM only recommends items that fall within a certain price range.
Although these features can be included in the prompt to trigger an LLM's generation, recommendations provided by the LLM may still fail to meet the user's requirements.
Current research on the controllability of LLM-based recommendation mainly focuses on controlling the explanations \cite{PETER,PEPLER,P5}, while more research is urgently needed on controlling recommendations generated by LLM.

\subsection{Inference Efficiency}

As an LLM contains a huge amount of parameters and RS is a latency-sensitive application, the efficiency of LLM-based recommendation models is vital.
The training efficiency can be improved by either option tuning \cite{M6} or adapter tuning \cite{VIP5}.
To reduce LLM's training time, \cite{POD} propose a task-alternated training strategy to deal with multiple recommendation tasks.
Since the training efficiency of LLM can be improved in an offline environment and usually an LLM does not have to be retrained too frequently, it is not as important as the inference efficiency problem.
\cite{M6} pre-compute the first few layers of an LLM and cache the results to improve its inference efficiency. However, this strategy may only apply to a specific LLM architecture that represents users and items with metadata.
\cite{POD} observe that LLM's inference time can be slightly reduced when the discrete prompt is removed.
In summary, there is still much room to further improve the inference efficiency of LLM-based recommendation models.

\subsection{Multimodal Recommendation}

In addition to text, data of other modalities can also be leveraged by LLM, as long as they can be represented as a sequence of tokens that can be integrated into textual sentences, as in the case of DALL·E \cite{DALLE} and Sora\footnote{\url{https://openai.com/sora}}.
For example, \cite{VIP5} incorporate item images into an LLM to improve its performance on recommendation tasks.
Regarding image generation, \cite{METER} generate visual explanations for recommendations based on a vision-language model, and \cite{M6} synthesize images for product design.
Besides images, videos and audios can also be generated in an auto-regressive way \cite{rubenstein2023audiopalm,VideoGPT}, which makes LLM-based multimodal recommendation a promising direction, such as short video recommendation and music recommendation.
Furthermore, when there is no available item that caters to a user's interest in the item repository, the system can create new items, especially for fashion recommendation, e.g., clothes.
Even if the generated items do not fully meet a user's requirements, they can be used to retrieve existing similar items or spark designers' creativity for improved design.
Meanwhile, model developers should guarantee the authenticity of machine-generated content to prevent users from having a negative experience, e.g., a picture of a Hawaiian attraction captioned South Korea for travel recommendation.

\subsection{Cold-start Recommendation}

As LLM have learned world knowledge during the pre-training stage, they can perform recommendation tasks even if they are not fine-tuned on recommendation-specific datasets.
A typical example is ChatGPT, which can be instructed to perform various recommendation tasks as discussed in the previous section \cite{liu2023chatgpt}.
The underlying reason is that users' preferences and items' attributes can be expressed in natural language.
As a result, LLM-based recommendation models have the potential to mitigate the well-known cold-start problem where there is limited or even no interaction regarding new users or items.
Although the interaction data are insufficient, we may still make use of their metadata for recommendation, such as user demographic information and item description information.

\section{Conclusions} \label{sec:conclude}
In this survey, we have reviewed the recent progress of LLM-based generative recommendation and provided a general formulation for each generative recommendation task according to relevant research.
To encourage researchers to explore this promising direction, we have elaborated on its advantages compared to traditional RS, generalized the definition of IDs, and summarized various ID creation methods.
We have also pointed out several prospects that might be worth in-depth exploration. 
We anticipate a future where LLM and RS will be nicely integrated to create high-quality personalized services in various scenarios.

\section{Acknowledgements}

This work was supported by Hong Kong Baptist University IG-FNRA project (RC-FNRA-IG/21-22/SCI/01) and Hong Kong Research Grants Council (RGC) Postdoctoral Fellowship Scheme (PDFS2223-2S02), and partially supported by NSF IIS-1910154, 2007907, and 2046457.
Any opinions, findings, conclusions or recommendations expressed in this material are those of the authors and do not necessarily reflect those of the sponsors.

\nocite{*}
\section{Bibliographical References}\label{sec:reference}

\bibliographystyle{lrec-coling2024-natbib}
\bibliography{bibliography}

\begin{thebibliography}{89}
\expandafter\ifx\csname natexlab\endcsname\relax\def\natexlab#1{#1}\fi

\bibitem[{Azamfirei et~al.(2023)Azamfirei, Kudchadkar, and
  Fackler}]{azamfirei2023large}
Razvan Azamfirei, Sapna~R Kudchadkar, and James Fackler. 2023.
\newblock Large language models and the perils of their hallucinations.
\newblock \emph{Critical Care}, 27(1):1--2.

\bibitem[{Bao et~al.(2023{\natexlab{a}})Bao, Zhang, Wang, Zhang, Yang, Luo,
  Feng, He, and Tian}]{BIGRec}
Keqin Bao, Jizhi Zhang, Wenjie Wang, Yang Zhang, Zhengyi Yang, Yancheng Luo,
  Fuli Feng, Xiangnaan He, and Qi~Tian. 2023{\natexlab{a}}.
\newblock A bi-step grounding paradigm for large language models in
  recommendation systems.
\newblock \emph{arXiv preprint arXiv:2308.08434}.

\bibitem[{Bao et~al.(2023{\natexlab{b}})Bao, Zhang, Zhang, Wang, Feng, and
  He}]{TALLRec}
Keqin Bao, Jizhi Zhang, Yang Zhang, Wenjie Wang, Fuli Feng, and Xiangnan He.
  2023{\natexlab{b}}.
\newblock Tallrec: An effective and efficient tuning framework to align large
  language model with recommendation.
\newblock In \emph{Proceedings of the 17th ACM Conference on Recommender
  Systems}.

\bibitem[{Carraro and Bridge(2024)}]{carraro2024enhancing}
Diego Carraro and Derek Bridge. 2024.
\newblock Enhancing recommendation diversity by re-ranking with large language
  models.
\newblock \emph{arXiv preprint arXiv:2401.11506}.

\bibitem[{Chen et~al.(2023)Chen, Liu, Huang, Wu, Liu, Jiang, Pu, Lei, Chen,
  Wang et~al.}]{chen2023large}
Jin Chen, Zheng Liu, Xu~Huang, Chenwang Wu, Qi~Liu, Gangwei Jiang, Yuanhao Pu,
  Yuxuan Lei, Xiaolong Chen, Xingmei Wang, et~al. 2023.
\newblock When large language models meet personalization: Perspectives of
  challenges and opportunities.
\newblock \emph{arXiv preprint arXiv:2307.16376}.

\bibitem[{Chen et~al.(2019)Chen, Chen, Xu, Zhang, Cao, Qin, and Zha}]{VECF}
Xu~Chen, Hanxiong Chen, Hongteng Xu, Yongfeng Zhang, Yixin Cao, Zheng Qin, and
  Hongyuan Zha. 2019.
\newblock Personalized fashion recommendation with visual explanations based on
  multimodal attention network: Towards visually explainable recommendation.
\newblock In \emph{Proceedings of the 42nd International ACM SIGIR Conference
  on Research and Development in Information Retrieval}, pages 765--774.

\bibitem[{Covington et~al.(2016)Covington, Adams, and
  Sargin}]{covington2016deep}
Paul Covington, Jay Adams, and Emre Sargin. 2016.
\newblock Deep neural networks for youtube recommendations.
\newblock In \emph{Proceedings of the 10th ACM conference on recommender
  systems}, pages 191--198.

\bibitem[{Cui et~al.(2022)Cui, Ma, Zhou, Zhou, and Yang}]{M6}
Zeyu Cui, Jianxin Ma, Chang Zhou, Jingren Zhou, and Hongxia Yang. 2022.
\newblock M6-rec: Generative pretrained language models are open-ended
  recommender systems.
\newblock \emph{arXiv preprint arXiv:2205.08084}.

\bibitem[{Dai et~al.(2023)Dai, Shao, Zhao, Yu, Si, Xu, Sun, Zhang, and
  Xu}]{dai2023uncovering}
Sunhao Dai, Ninglu Shao, Haiyuan Zhao, Weijie Yu, Zihua Si, Chen Xu, Zhongxiang
  Sun, Xiao Zhang, and Jun Xu. 2023.
\newblock Uncovering chatgpt's capabilities in recommender systems.
\newblock In \emph{Proceedings of the 17th ACM Conference on Recommender
  Systems}.

\bibitem[{Di~Palma et~al.(2023)Di~Palma, Biancofiore, Anelli, Narducci,
  Di~Noia, and Di~Sciascio}]{di2023evaluating}
Dario Di~Palma, Giovanni~Maria Biancofiore, Vito~Walter Anelli, Fedelucio
  Narducci, Tommaso Di~Noia, and Eugenio Di~Sciascio. 2023.
\newblock Evaluating chatgpt as a recommender system: A rigorous approach.
\newblock \emph{arXiv preprint arXiv:2309.03613}.

\bibitem[{Fan et~al.(2023)Fan, Zhao, Li, Liu, Mei, Wang, Tang, and
  Li}]{fan2023recommender}
Wenqi Fan, Zihuai Zhao, Jiatong Li, Yunqing Liu, Xiaowei Mei, Yiqi Wang,
  Jiliang Tang, and Qing Li. 2023.
\newblock Recommender systems in the era of large language models (llms).
\newblock \emph{arXiv preprint arXiv:2307.02046}.

\bibitem[{Friedman et~al.(2023)Friedman, Ahuja, Allen, Tan, Sidahmed, Long,
  Xie, Schubiner, Patel, Lara et~al.}]{RecLLM}
Luke Friedman, Sameer Ahuja, David Allen, Terry Tan, Hakim Sidahmed, Changbo
  Long, Jun Xie, Gabriel Schubiner, Ajay Patel, Harsh Lara, et~al. 2023.
\newblock Leveraging large language models in conversational recommender
  systems.
\newblock \emph{arXiv preprint arXiv:2305.07961}.

\bibitem[{Ge et~al.(2024)Ge, Hua, Mei, Tan, Xu, Li, Zhang
  et~al.}]{ge2024openagi}
Yingqiang Ge, Wenyue Hua, Kai Mei, Juntao Tan, Shuyuan Xu, Zelong Li, Yongfeng
  Zhang, et~al. 2024.
\newblock Openagi: When llm meets domain experts.
\newblock In \emph{Advances in Neural Information Processing Systems}.

\bibitem[{Geng et~al.(2022{\natexlab{a}})Geng, Fu, Ge, Li, De~Melo, and
  Zhang}]{METER}
Shijie Geng, Zuohui Fu, Yingqiang Ge, Lei Li, Gerard De~Melo, and Yongfeng
  Zhang. 2022{\natexlab{a}}.
\newblock Improving personalized explanation generation through visualization.
\newblock In \emph{Proceedings of the 60th Annual Meeting of the Association
  for Computational Linguistics (Volume 1: Long Papers)}, pages 244--255.

\bibitem[{Geng et~al.(2022{\natexlab{b}})Geng, Fu, Tan, Ge, De~Melo, and
  Zhang}]{geng2022path}
Shijie Geng, Zuohui Fu, Juntao Tan, Yingqiang Ge, Gerard De~Melo, and Yongfeng
  Zhang. 2022{\natexlab{b}}.
\newblock Path language modeling over knowledge graphsfor explainable
  recommendation.
\newblock In \emph{Proceedings of the ACM Web Conference 2022}, pages 946--955.

\bibitem[{Geng et~al.(2022{\natexlab{c}})Geng, Liu, Fu, Ge, and Zhang}]{P5}
Shijie Geng, Shuchang Liu, Zuohui Fu, Yingqiang Ge, and Yongfeng Zhang.
  2022{\natexlab{c}}.
\newblock Recommendation as language processing (rlp): A unified pretrain,
  personalized prompt \& predict paradigm (p5).
\newblock In \emph{Sixteenth ACM Conference on Recommender Systems}.

\bibitem[{Geng et~al.(2023)Geng, Tan, Liu, Fu, and Zhang}]{VIP5}
Shijie Geng, Juntao Tan, Shuchang Liu, Zuohui Fu, and Yongfeng Zhang. 2023.
\newblock Vip5: Towards multimodal foundation models for recommendation.
\newblock In \emph{Proceedings of the 2023 Conference on Empirical Methods in
  Natural Language Processing}.

\bibitem[{He et~al.(2023)He, Xie, Jha, Steck, Liang, Feng, Majumder, Kallus,
  and McAuley}]{he2023large}
Zhankui He, Zhouhang Xie, Rahul Jha, Harald Steck, Dawen Liang, Yesu Feng,
  Bodhisattwa~Prasad Majumder, Nathan Kallus, and Julian McAuley. 2023.
\newblock Large language models as zero-shot conversational recommenders.
\newblock In \emph{Proceedings of the 32nd ACM International Conference on
  Information \& Knowledge Management}.

\bibitem[{Hou et~al.(2024)Hou, Zhang, Lin, Lu, Xie, McAuley, and
  Zhao}]{hou2023large}
Yupeng Hou, Junjie Zhang, Zihan Lin, Hongyu Lu, Ruobing Xie, Julian McAuley,
  and Wayne~Xin Zhao. 2024.
\newblock Large language models are zero-shot rankers for recommender systems.
\newblock In \emph{46th European Conference on Information Retrieval}.

\bibitem[{Hua et~al.(2023{\natexlab{a}})Hua, Fan, Li, Mei, Ji, Ge, Hemphill,
  and Zhang}]{hua2023war}
Wenyue Hua, Lizhou Fan, Lingyao Li, Kai Mei, Jianchao Ji, Yingqiang Ge, Libby
  Hemphill, and Yongfeng Zhang. 2023{\natexlab{a}}.
\newblock War and peace (waragent): Large language model-based multi-agent
  simulation of world wars.
\newblock \emph{arXiv preprint arXiv:2311.17227}.

\bibitem[{Hua et~al.(2024)Hua, Ge, Xu, Ji, and Zhang}]{UP5}
Wenyue Hua, Yingqiang Ge, Shuyuan Xu, Jianchao Ji, and Yongfeng Zhang. 2024.
\newblock Up5: Unbiased foundation model for fairness-aware recommendation.
\newblock In \emph{18th Conference of the European Chapter of the Association
  for Computational Linguistics}.

\bibitem[{Hua et~al.(2023{\natexlab{b}})Hua, Xu, Ge, and Zhang}]{Index}
Wenyue Hua, Shuyuan Xu, Yingqiang Ge, and Yongfeng Zhang. 2023{\natexlab{b}}.
\newblock How to index item ids for recommendation foundation models.
\newblock In \emph{Proceedings of 1st International ACM SIGIR Conference on
  Information Retrieval in the Asia Pacific}.

\bibitem[{Huang et~al.(2024)Huang, Yu, Xie, Zhang, Yao, and
  McAuley}]{huang2024foundation}
Chengkai Huang, Tong Yu, Kaige Xie, Shuai Zhang, Lina Yao, and Julian McAuley.
  2024.
\newblock Foundation models for recommender systems: A survey and new
  perspectives.
\newblock \emph{arXiv preprint arXiv:2402.11143}.

\bibitem[{Huang et~al.(2023)Huang, Lian, Lei, Yao, Lian, and
  Xie}]{InteRecAgent}
Xu~Huang, Jianxun Lian, Yuxuan Lei, Jing Yao, Defu Lian, and Xing Xie. 2023.
\newblock Recommender ai agent: Integrating large language models for
  interactive recommendations.
\newblock \emph{arXiv preprint arXiv:2308.16505}.

\bibitem[{Jannach et~al.(2021)Jannach, Manzoor, Cai, and Chen}]{Conversation}
Dietmar Jannach, Ahtsham Manzoor, Wanling Cai, and Li~Chen. 2021.
\newblock A survey on conversational recommender systems.
\newblock \emph{ACM Computing Surveys (CSUR)}, 54(5):1--36.

\bibitem[{Ji et~al.(2024)Ji, Li, Xu, Hua, Ge, Tan, and Zhang}]{GenRec}
Jianchao Ji, Zelong Li, Shuyuan Xu, Wenyue Hua, Yingqiang Ge, Juntao Tan, and
  Yongfeng Zhang. 2024.
\newblock Genrec: Large language model for generative recommendation.
\newblock In \emph{46th European Conference on Information Retrieval}.

\bibitem[{Li et~al.(2020)Li, Zhang, and Chen}]{NETE}
Lei Li, Yongfeng Zhang, and Li~Chen. 2020.
\newblock Generate neural template explanations for recommendation.
\newblock In \emph{Proceedings of the 29th ACM International Conference on
  Information \& Knowledge Management}, pages 755--764.

\bibitem[{Li et~al.(2021)Li, Zhang, and Chen}]{PETER}
Lei Li, Yongfeng Zhang, and Li~Chen. 2021.
\newblock Personalized transformer for explainable recommendation.
\newblock In \emph{Proceedings of the 59th Annual Meeting of the Association
  for Computational Linguistics}, pages 4947--4957.

\bibitem[{Li et~al.(2023{\natexlab{a}})Li, Zhang, and Chen}]{PEPLER}
Lei Li, Yongfeng Zhang, and Li~Chen. 2023{\natexlab{a}}.
\newblock Personalized prompt learning for explainable recommendation.
\newblock \emph{ACM Transactions on Information Systems}, 41(4):1--26.

\bibitem[{Li et~al.(2023{\natexlab{b}})Li, Zhang, and Chen}]{POD}
Lei Li, Yongfeng Zhang, and Li~Chen. 2023{\natexlab{b}}.
\newblock Prompt distillation for efficient llm-based recommendation.
\newblock In \emph{Proceedings of the 32nd ACM International Conference on
  Information \& Knowledge Management}.

\bibitem[{Li and Tuzhilin(2019)}]{Review}
Pan Li and Alexander Tuzhilin. 2019.
\newblock Towards controllable and personalized review generation.
\newblock In \emph{Proceedings of the 2019 Conference on Empirical Methods in
  Natural Language Processing and the 9th International Joint Conference on
  Natural Language Processing (EMNLP-IJCNLP)}, pages 3237--3245, Hong Kong,
  China. Association for Computational Linguistics.

\bibitem[{Li et~al.(2023{\natexlab{c}})Li, Deng, Cheng, Yuan, Zhang, and
  Yuan}]{li2023exploring}
Ruyu Li, Wenhao Deng, Yu~Cheng, Zheng Yuan, Jiaqi Zhang, and Fajie Yuan.
  2023{\natexlab{c}}.
\newblock Exploring the upper limits of text-based collaborative filtering
  using large language models: Discoveries and insights.
\newblock \emph{arXiv preprint arXiv:2305.11700}.

\bibitem[{Li et~al.(2023{\natexlab{d}})Li, Zhang, and
  Malthouse}]{li2023exploringnews}
Xinyi Li, Yongfeng Zhang, and Edward~C Malthouse. 2023{\natexlab{d}}.
\newblock Exploring fine-tuning chatgpt for news recommendation.
\newblock \emph{arXiv preprint arXiv:2311.05850}.

\bibitem[{Li et~al.(2023{\natexlab{e}})Li, Zhang, and Malthouse}]{PBNR}
Xinyi Li, Yongfeng Zhang, and Edward~C Malthouse. 2023{\natexlab{e}}.
\newblock Pbnr: Prompt-based news recommender system.
\newblock \emph{arXiv preprint arXiv:2304.07862}.

\bibitem[{Li et~al.(2023{\natexlab{f}})Li, Zhang, and Malthouse}]{News}
Xinyi Li, Yongfeng Zhang, and Edward~C Malthouse. 2023{\natexlab{f}}.
\newblock A preliminary study of chatgpt on news recommendation:
  Personalization, provider fairness, fake news.
\newblock In \emph{11th International Workshop on News Recommendation and
  Analytics in conjunction with ACM RecSys 2023}.

\bibitem[{Li et~al.(2023{\natexlab{g}})Li, Chen, Zhang, and Liang}]{BookGPT}
Zhiyu Li, Yanfang Chen, Xuan Zhang, and Xun Liang. 2023{\natexlab{g}}.
\newblock \href {https://doi.org/10.3390/electronics12224654} {Bookgpt: A
  general framework for book recommendation empowered by large language model}.
\newblock \emph{Electronics}, 12(22).

\bibitem[{Liao et~al.(2023)Liao, Li, Yang, Wu, Yuan, Wang, and He}]{LLaRa}
Jiayi Liao, Sihang Li, Zhengyi Yang, Jiancan Wu, Yancheng Yuan, Xiang Wang, and
  Xiangnan He. 2023.
\newblock Llara: Aligning large language models with sequential recommenders.
\newblock \emph{arXiv preprint arXiv:2312.02445}.

\bibitem[{Lin(2004)}]{ROUGE}
Chin-Yew Lin. 2004.
\newblock Rouge: A package for automatic evaluation of summaries.
\newblock In \emph{Text summarization branches out}, pages 74--81.

\bibitem[{Lin and Zhang(2023)}]{lin2023sparks}
Guo Lin and Yongfeng Zhang. 2023.
\newblock Sparks of artificial general recommender (agr): Early experiments
  with chatgpt.
\newblock \emph{Algorithms}.

\bibitem[{Lin et~al.(2023{\natexlab{a}})Lin, Dai, Xi, Liu, Chen, Li, Zhu, Guo,
  Yu, Tang et~al.}]{Industry}
Jianghao Lin, Xinyi Dai, Yunjia Xi, Weiwen Liu, Bo~Chen, Xiangyang Li, Chenxu
  Zhu, Huifeng Guo, Yong Yu, Ruiming Tang, et~al. 2023{\natexlab{a}}.
\newblock How can recommender systems benefit from large language models: A
  survey.
\newblock \emph{arXiv preprint arXiv:2306.05817}.

\bibitem[{Lin et~al.(2024)Lin, Shan, Zhu, Du, Chen, Quan, Tang, Yu, and
  Zhang}]{ReLLa}
Jianghao Lin, Rong Shan, Chenxu Zhu, Kounianhua Du, Bo~Chen, Shigang Quan,
  Ruiming Tang, Yong Yu, and Weinan Zhang. 2024.
\newblock Rella: Retrieval-enhanced large language models for lifelong
  sequential behavior comprehension in recommendation.
\newblock In \emph{The Web Conference 2024}.

\bibitem[{Lin et~al.(2023{\natexlab{b}})Lin, Wang, Li, Feng, Ng, and
  Chua}]{TransRec}
Xinyu Lin, Wenjie Wang, Yongqi Li, Fuli Feng, See-Kiong Ng, and Tat-Seng Chua.
  2023{\natexlab{b}}.
\newblock A multi-facet paradigm to bridge large language model and
  recommendation.
\newblock \emph{arXiv preprint arXiv:2310.06491}.

\bibitem[{Liu et~al.(2023{\natexlab{a}})Liu, Liu, Lv, Zhou, and
  Zhang}]{liu2023chatgpt}
Junling Liu, Chao Liu, Renjie Lv, Kang Zhou, and Yan Zhang. 2023{\natexlab{a}}.
\newblock Is chatgpt a good recommender? a preliminary study.
\newblock In \emph{1st Workshop on Recommendation with Generative Models
  (GenRec) at CIKM 2023}.

\bibitem[{Liu et~al.(2023{\natexlab{b}})Liu, Liu, Zhou, Ye, Chong, Zhou, Xie,
  Cao, Wang, You et~al.}]{LLMRec}
Junling Liu, Chao Liu, Peilin Zhou, Qichen Ye, Dading Chong, Kang Zhou, Yueqi
  Xie, Yuwei Cao, Shoujin Wang, Chenyu You, et~al. 2023{\natexlab{b}}.
\newblock Llmrec: Benchmarking large language models on recommendation task.
\newblock \emph{arXiv preprint arXiv:2308.12241}.

\bibitem[{Liu et~al.(2023{\natexlab{c}})Liu, Zhang, and Gulla}]{training}
Peng Liu, Lemei Zhang, and Jon~Atle Gulla. 2023{\natexlab{c}}.
\newblock Pre-train, prompt and recommendation: A comprehensive survey of
  language modelling paradigm adaptations in recommender systems.
\newblock \emph{Transactions of the Association for Computational Linguistics}.

\bibitem[{Liu et~al.(2023{\natexlab{d}})Liu, Yuan, Fu, Jiang, Hayashi, and
  Neubig}]{Prompt}
Pengfei Liu, Weizhe Yuan, Jinlan Fu, Zhengbao Jiang, Hiroaki Hayashi, and
  Graham Neubig. 2023{\natexlab{d}}.
\newblock Pre-train, prompt, and predict: A systematic survey of prompting
  methods in natural language processing.
\newblock \emph{ACM Computing Surveys}, 55(9):1--35.

\bibitem[{Luettin et~al.(2019)Luettin, Rothermel, and Andrew}]{vehicle}
Juergen Luettin, Susanne Rothermel, and Mark Andrew. 2019.
\newblock \href {https://doi.org/10.1145/3298689.3346958} {Future of in-vehicle
  recommendation systems @ bosch}.
\newblock In \emph{Proceedings of the 13th ACM Conference on Recommender
  Systems}, RecSys '19, page 524, New York, NY, USA. Association for Computing
  Machinery.

\bibitem[{Luo et~al.(2023)Luo, He, Zhao, Huang, Zhou, Li, Xiao, Zhan, and
  Song}]{RecRanker}
Sichun Luo, Bowei He, Haohan Zhao, Yinya Huang, Aojun Zhou, Zongpeng Li,
  Yuanzhang Xiao, Mingjie Zhan, and Linqi Song. 2023.
\newblock Recranker: Instruction tuning large language model as ranker for
  top-k recommendation.
\newblock \emph{arXiv preprint arXiv:2312.16018}.

\bibitem[{Luo et~al.(2024)Luo, Yao, He, Huang, Zhou, Zhang, Xiao, Zhan, and
  Song}]{Llama4Rec}
Sichun Luo, Yuxuan Yao, Bowei He, Yinya Huang, Aojun Zhou, Xinyi Zhang,
  Yuanzhang Xiao, Mingjie Zhan, and Linqi Song. 2024.
\newblock Integrating large language models into recommendation via mutual
  augmentation and adaptive aggregation.
\newblock \emph{arXiv preprint arXiv:2401.13870}.

\bibitem[{Mialon et~al.(2023)Mialon, Dess{\`\i}, Lomeli, Nalmpantis, Pasunuru,
  Raileanu, Rozi{\`e}re, Schick, Dwivedi-Yu, Celikyilmaz
  et~al.}]{mialon2023augmented}
Gr{\'e}goire Mialon, Roberto Dess{\`\i}, Maria Lomeli, Christoforos Nalmpantis,
  Ram Pasunuru, Roberta Raileanu, Baptiste Rozi{\`e}re, Timo Schick, Jane
  Dwivedi-Yu, Asli Celikyilmaz, et~al. 2023.
\newblock Augmented language models: a survey.
\newblock \emph{Transactions on Machine Learning Research}.

\bibitem[{Papineni et~al.(2002)Papineni, Roukos, Ward, and Zhu}]{BLEU}
Kishore Papineni, Salim Roukos, Todd Ward, and Wei-Jing Zhu. 2002.
\newblock Bleu: a method for automatic evaluation of machine translation.
\newblock In \emph{Proceedings of the 40th annual meeting of the Association
  for Computational Linguistics}, pages 311--318.

\bibitem[{Park et~al.(2023)Park, O'Brien, Cai, Morris, Liang, and
  Bernstein}]{agent}
Joon~Sung Park, Joseph O'Brien, Carrie~Jun Cai, Meredith~Ringel Morris, Percy
  Liang, and Michael~S. Bernstein. 2023.
\newblock \href {https://doi.org/10.1145/3586183.3606763} {Generative agents:
  Interactive simulacra of human behavior}.
\newblock In \emph{Proceedings of the 36th Annual ACM Symposium on User
  Interface Software and Technology}, UIST '23, New York, NY, USA. Association
  for Computing Machinery.

\bibitem[{Parra and Brusilovsky(2015)}]{parra2015user}
Denis Parra and Peter Brusilovsky. 2015.
\newblock User-controllable personalization: A case study with setfusion.
\newblock \emph{International Journal of Human-Computer Studies}, 78:43--67.

\bibitem[{Petrov and Macdonald(2023)}]{SVD}
Aleksandr~V Petrov and Craig Macdonald. 2023.
\newblock Generative sequential recommendation with gptrec.
\newblock In \emph{Gen-IR@SIGIR 2023: The First Workshop on Generative
  Information Retrieval}.

\bibitem[{Rahdari et~al.(2024)Rahdari, Ding, Fan, Ma, Chen, Deoras, and
  Kveton}]{Logic-Scaffolding}
Behnam Rahdari, Hao Ding, Ziwei Fan, Yifei Ma, Zhuotong Chen, Anoop Deoras, and
  Branislav Kveton. 2024.
\newblock Logic-scaffolding: Personalized aspect-instructed recommendation
  explanation generation using llms.
\newblock In \emph{Proceedings of the Seventeenth ACM International Conference
  on Web Search and Data Mining}.

\bibitem[{Ramesh et~al.(2021)Ramesh, Pavlov, Goh, Gray, Voss, Radford, Chen,
  and Sutskever}]{DALLE}
Aditya Ramesh, Mikhail Pavlov, Gabriel Goh, Scott Gray, Chelsea Voss, Alec
  Radford, Mark Chen, and Ilya Sutskever. 2021.
\newblock Zero-shot text-to-image generation.
\newblock In \emph{International Conference on Machine Learning}, pages
  8821--8831. PMLR.

\bibitem[{Ravaut et~al.(2024)Ravaut, Zhang, Xu, Sun, and Liu}]{PECRS}
Mathieu Ravaut, Hao Zhang, Lu~Xu, Aixin Sun, and Yong Liu. 2024.
\newblock Parameter-efficient conversational recommender system as a language
  processing task.
\newblock In \emph{18th Conference of the European Chapter of the Association
  for Computational Linguistics}.

\bibitem[{Rubenstein et~al.(2023)Rubenstein, Asawaroengchai, Nguyen, Bapna,
  Borsos, Quitry, Chen, Badawy, Han, Kharitonov
  et~al.}]{rubenstein2023audiopalm}
Paul~K Rubenstein, Chulayuth Asawaroengchai, Duc~Dung Nguyen, Ankur Bapna,
  Zal{\'a}n Borsos, F{\'e}lix de~Chaumont Quitry, Peter Chen, Dalia~El Badawy,
  Wei Han, Eugene Kharitonov, et~al. 2023.
\newblock Audiopalm: A large language model that can speak and listen.
\newblock \emph{arXiv preprint arXiv:2306.12925}.

\bibitem[{Schafer et~al.(2002)Schafer, Konstan, and Riedl}]{schafer2002meta}
J~Ben Schafer, Joseph~A Konstan, and John Riedl. 2002.
\newblock Meta-recommendation systems: user-controlled integration of diverse
  recommendations.
\newblock In \emph{Proceedings of the eleventh international conference on
  Information and knowledge management}, pages 43--51.

\bibitem[{SCIENCE(1977)}]{beam}
CARNEGIE-MELLON UNIV PITTSBURGH PA DEPT OF~COMPUTER SCIENCE. 1977.
\newblock \emph{Speech Understanding Systems. Summary of Results of the
  Five-Year Research Effort at Carnegie-Mellon University}.

\bibitem[{Sun and Zhang(2018)}]{sun2018conversational}
Yueming Sun and Yi~Zhang. 2018.
\newblock Conversational recommender system.
\newblock In \emph{The 41st international acm sigir conference on research \&
  development in information retrieval}, pages 235--244.

\bibitem[{Tan et~al.(2023)Tan, Ge, Zhu, Xia, Luo, Ji, and Zhang}]{tan2023user}
Juntao Tan, Yingqiang Ge, Yan Zhu, Yinglong Xia, Jiebo Luo, Jianchao Ji, and
  Yongfeng Zhang. 2023.
\newblock User-controllable recommendation via counterfactual retrospective and
  prospective explanations.
\newblock In \emph{26th European Conference on Artificial Intelligence}.

\bibitem[{Vaswani et~al.(2017)Vaswani, Shazeer, Parmar, Uszkoreit, Jones,
  Gomez, Kaiser, and Polosukhin}]{Transformer}
Ashish Vaswani, Noam Shazeer, Niki Parmar, Jakob Uszkoreit, Llion Jones,
  Aidan~N Gomez, {\L}ukasz Kaiser, and Illia Polosukhin. 2017.
\newblock Attention is all you need.
\newblock In \emph{Advances in neural information processing systems}.

\bibitem[{Vats et~al.(2024)Vats, Jain, Raja, and Chadha}]{vats2024exploring}
Arpita Vats, Vinija Jain, Rahul Raja, and Aman Chadha. 2024.
\newblock Exploring the impact of large language models on recommender systems:
  An extensive review.
\newblock \emph{arXiv preprint arXiv:2402.18590}.

\bibitem[{Von~Luxburg(2007)}]{Spectral}
Ulrike Von~Luxburg. 2007.
\newblock A tutorial on spectral clustering.
\newblock \emph{Statistics and computing}, 17:395--416.

\bibitem[{Wang and Lim(2023)}]{NIR}
Lei Wang and Ee-Peng Lim. 2023.
\newblock Zero-shot next-item recommendation using large pretrained language
  models.
\newblock \emph{arXiv preprint arXiv:2304.03153}.

\bibitem[{Wang et~al.(2023{\natexlab{a}})Wang, Zhang, Yang, Chen, Tang, Zhang,
  Chen, Lin, Song, Zhao, Xu, Dou, Wang, and Wen}]{RecAgent}
Lei Wang, Jingsen Zhang, Hao Yang, Zhiyuan Chen, Jiakai Tang, Zeyu Zhang,
  Xu~Chen, Yankai Lin, Ruihua Song, Wayne~Xin Zhao, Jun Xu, Zhicheng Dou, Jun
  Wang, and Ji-Rong Wen. 2023{\natexlab{a}}.
\newblock User behavior simulation with large language model based agents.
\newblock \emph{arXiv preprint arXiv:2306.02552}.

\bibitem[{Wang et~al.(2023{\natexlab{b}})Wang, Wang, Wang, Sanjabi, Liu,
  Firooz, Wang, and Nie}]{COFFEE}
Nan Wang, Qifan Wang, Yi-Chia Wang, Maziar Sanjabi, Jingzhou Liu, Hamed Firooz,
  Hongning Wang, and Shaoliang Nie. 2023{\natexlab{b}}.
\newblock \href {https://doi.org/10.18653/v1/2023.emnlp-main.819} {{COFFEE}:
  Counterfactual fairness for personalized text generation in explainable
  recommendation}.
\newblock In \emph{Proceedings of the 2023 Conference on Empirical Methods in
  Natural Language Processing}, pages 13258--13275, Singapore. Association for
  Computational Linguistics.

\bibitem[{Wang et~al.(2022)Wang, Feng, Nie, and Chua}]{wang2022user}
Wenjie Wang, Fuli Feng, Liqiang Nie, and Tat-Seng Chua. 2022.
\newblock User-controllable recommendation against filter bubbles.
\newblock In \emph{Proceedings of the 45th International ACM SIGIR Conference
  on Research and Development in Information Retrieval}, pages 1251--1261.

\bibitem[{Wang et~al.(2023{\natexlab{c}})Wang, Tang, Zhao, Wang, and
  Wen}]{wang2023rethinking}
Xiaolei Wang, Xinyu Tang, Wayne~Xin Zhao, Jingyuan Wang, and Ji-Rong Wen.
  2023{\natexlab{c}}.
\newblock Rethinking the evaluation for conversational recommendation in the
  era of large language models.
\newblock In \emph{Proceedings of the 2023 Conference on Empirical Methods in
  Natural Language Processing}.

\bibitem[{Wang et~al.(2023{\natexlab{d}})Wang, Jiang, Chen, Yang, Zhou, Cho,
  Fan, Huang, Lu, and Yang}]{RecMind}
Yancheng Wang, Ziyan Jiang, Zheng Chen, Fan Yang, Yingxue Zhou, Eunah Cho, Xing
  Fan, Xiaojiang Huang, Yanbin Lu, and Yingzhen Yang. 2023{\natexlab{d}}.
\newblock Recmind: Large language model powered agent for recommendation.
\newblock \emph{arXiv preprint arXiv:2308.14296}.

\bibitem[{Wu et~al.(2024)Wu, Qiu, Zheng, Zhu, and Chen}]{wu2023exploring}
Likang Wu, Zhaopeng Qiu, Zhi Zheng, Hengshu Zhu, and Enhong Chen. 2024.
\newblock Exploring large language model for graph data understanding in online
  job recommendations.
\newblock In \emph{Proceedings of the AAAI Conference on Artificial
  Intelligence}.

\bibitem[{Wu et~al.(2023)Wu, Zheng, Qiu, Wang, Gu, Shen, Qin, Zhu, Zhu, Liu
  et~al.}]{wu2023survey}
Likang Wu, Zhi Zheng, Zhaopeng Qiu, Hao Wang, Hongchao Gu, Tingjia Shen, Chuan
  Qin, Chen Zhu, Hengshu Zhu, Qi~Liu, et~al. 2023.
\newblock A survey on large language models for recommendation.
\newblock \emph{arXiv preprint arXiv:2305.19860}.

\bibitem[{Xu et~al.(2023{\natexlab{a}})Xu, Wang, Li, Pang, Xu, and
  Chua}]{xu2023llms}
Chen Xu, Wenjie Wang, Yuxin Li, Liang Pang, Jun Xu, and Tat-Seng Chua.
  2023{\natexlab{a}}.
\newblock Do llms implicitly exhibit user discrimination in recommendation? an
  empirical study.
\newblock \emph{arXiv preprint arXiv:2311.07054}.

\bibitem[{Xu et~al.(2023{\natexlab{b}})Xu, Hua, and Zhang}]{OpenP5}
Shuyuan Xu, Wenyue Hua, and Yongfeng Zhang. 2023{\natexlab{b}}.
\newblock Openp5: Benchmarking foundation models for recommendation.
\newblock \emph{arXiv preprint arXiv:2306.11134}.

\bibitem[{Yan et~al.(2021)Yan, Zhang, Abbeel, and Srinivas}]{VideoGPT}
Wilson Yan, Yunzhi Zhang, Pieter Abbeel, and Aravind Srinivas. 2021.
\newblock Videogpt: Video generation using vq-vae and transformers.
\newblock \emph{arXiv preprint arXiv:2104.10157}.

\bibitem[{Yang et~al.(2023)Yang, Chen, Jiang, Cho, Huang, and Lu}]{PALR}
Fan Yang, Zheng Chen, Ziyan Jiang, Eunah Cho, Xiaojiang Huang, and Yanbin Lu.
  2023.
\newblock Palr: Personalization aware llms for recommendation.
\newblock In \emph{Gen-IR@SIGIR 2023: The First Workshop on Generative
  Information Retrieval}.

\bibitem[{Ye et~al.(2024)Ye, Zhang, Wang, Xu, and Zhang}]{ye2023natural}
Ruosong Ye, Caiqi Zhang, Runhui Wang, Shuyuan Xu, and Yongfeng Zhang. 2024.
\newblock Language is all a graph needs.
\newblock In \emph{18th Conference of the European Chapter of the Association
  for Computational Linguistics}.

\bibitem[{Yu et~al.(2023)Yu, Li, Wang, Li, and Xu}]{yu2023counterfactual}
Dianer Yu, Qian Li, Xiangmeng Wang, Qing Li, and Guandong Xu. 2023.
\newblock Counterfactual explainable conversational recommendation.
\newblock \emph{IEEE Transactions on Knowledge and Data Engineering}.

\bibitem[{Zeghidour et~al.(2021)Zeghidour, Luebs, Omran, Skoglund, and
  Tagliasacchi}]{RQVAE}
Neil Zeghidour, Alejandro Luebs, Ahmed Omran, Jan Skoglund, and Marco
  Tagliasacchi. 2021.
\newblock Soundstream: An end-to-end neural audio codec.
\newblock \emph{IEEE/ACM Transactions on Audio, Speech, and Language
  Processing}, 30:495--507.

\bibitem[{Zhang et~al.(2023{\natexlab{a}})Zhang, Sheng, Chen, Li, Deng, Wang,
  and Chua}]{Agent4Rec}
An~Zhang, Leheng Sheng, Yuxin Chen, Hao Li, Yang Deng, Xiang Wang, and Tat-Seng
  Chua. 2023{\natexlab{a}}.
\newblock On generative agents in recommendation.
\newblock \emph{arXiv preprint arXiv:2310.10108}.

\bibitem[{Zhang et~al.(2023{\natexlab{b}})Zhang, Bao, Zhang, Wang, Feng, and
  He}]{zhang2023chatgpt}
Jizhi Zhang, Keqin Bao, Yang Zhang, Wenjie Wang, Fuli Feng, and Xiangnan He.
  2023{\natexlab{b}}.
\newblock Is chatgpt fair for recommendation? evaluating fairness in large
  language model recommendation.
\newblock In \emph{Proceedings of the 17th ACM Conference on Recommender
  Systems}.

\bibitem[{Zhang et~al.(2023{\natexlab{c}})Zhang, Xie, Hou, Zhao, Lin, and
  Wen}]{zhang2023recommendation}
Junjie Zhang, Ruobing Xie, Yupeng Hou, Wayne~Xin Zhao, Leyu Lin, and Ji-Rong
  Wen. 2023{\natexlab{c}}.
\newblock Recommendation as instruction following: A large language model
  empowered recommendation approach.
\newblock \emph{arXiv preprint arXiv:2305.07001}.

\bibitem[{Zhang et~al.(2020{\natexlab{a}})Zhang, Kishore, Wu, Weinberger, and
  Artzi}]{zhang2019bertscore}
Tianyi Zhang, Varsha Kishore, Felix Wu, Kilian~Q Weinberger, and Yoav Artzi.
  2020{\natexlab{a}}.
\newblock Bertscore: Evaluating text generation with bert.
\newblock In \emph{International Conference on Learning Representations
  (ICLR)}.

\bibitem[{Zhang et~al.(2018)Zhang, Chen, Ai, Yang, and
  Croft}]{zhang2018towards}
Yongfeng Zhang, Xu~Chen, Qingyao Ai, Liu Yang, and W~Bruce Croft. 2018.
\newblock Towards conversational search and recommendation: System ask, user
  respond.
\newblock In \emph{Proceedings of the 27th acm international conference on
  information and knowledge management}, pages 177--186.

\bibitem[{Zhang et~al.(2020{\natexlab{b}})Zhang, Chen et~al.}]{Explanation}
Yongfeng Zhang, Xu~Chen, et~al. 2020{\natexlab{b}}.
\newblock Explainable recommendation: A survey and new perspectives.
\newblock \emph{Foundations and Trends{\textregistered} in Information
  Retrieval}, 14(1):1--101.

\bibitem[{Zhang et~al.(2014)Zhang, Lai, Zhang, Zhang, Liu, and Ma}]{EFM}
Yongfeng Zhang, Guokun Lai, Min Zhang, Yi~Zhang, Yiqun Liu, and Shaoping Ma.
  2014.
\newblock Explicit factor models for explainable recommendation based on
  phrase-level sentiment analysis.
\newblock In \emph{Proceedings of the 37th international ACM SIGIR conference
  on Research \& development in information retrieval}, pages 83--92.

\bibitem[{Zhang et~al.(2021)Zhang, Ding, Shui, Ma, Zou, Deoras, and
  Wang}]{LMRecSys}
Yuhui Zhang, Hao Ding, Zeren Shui, Yifei Ma, James Zou, Anoop Deoras, and Hao
  Wang. 2021.
\newblock Language models as recommender systems: Evaluations and limitations.
\newblock In \emph{NeurIPS 2021 Workshop ICBINB}.

\bibitem[{Zheng et~al.(2023)Zheng, Hou, Lu, Chen, Zhao, and Wen}]{LC-Rec}
Bowen Zheng, Yupeng Hou, Hongyu Lu, Yu~Chen, Wayne~Xin Zhao, and Ji-Rong Wen.
  2023.
\newblock Adapting large language models by integrating collaborative semantics
  for recommendation.
\newblock \emph{arXiv preprint arXiv:2311.09049}.

\end{thebibliography}

\end{document}